\documentclass[%
  manuscript,%
  groupedaddress,%
  showpacs,%
  prl,%
  letterpaper,%
  amsfonts,%
  floatfix,%
]{revtex4}

\usepackage{amsmath,amssymb,bm}
\usepackage[pdftex]{graphicx}
\usepackage{subfigure}
\usepackage{hyperref}
\hypersetup{%
  breaklinks = {true},
  citecolor = {blue},
  colorlinks = {true},
  linkcolor = {red},
  pdfauthor = {\textcopyright\ Vitor M. Pereira},
  pdfcreator = {\LaTeX\ and \flqq hyperref\frqq},
  pdffitwindow = {true},
  pdfmenubar = {true},
  pdfpagelayout = {SinglePage},
  pdfstartview = {Fit},
  pdftoolbar = {true},
  plainpages = {false},
}

\renewcommand{\section}[1]{\par\emph{#1} -- }
\newcommand{\vF}{\ensuremath{v_\text{F\,}}}

\newcommand{\bn}{\ensuremath{\bm{n}}}

\newcommand{\br}{\ensuremath{\bm{r}}}
\newcommand{\bR}{\ensuremath{\bm{R}}}
\newcommand{\bu}{\ensuremath{\bm{u}}}
\newcommand{\dt}{\ensuremath{\delta t}}

\newcommand{\Fref}[1]{Fig.~\ref{#1}}
\newcommand{\Eqref}[1]{Eq.~\eqref{#1}}

\begin{document}


\title{All-graphene integrated circuits via strain engineering}


%

\author{Vitor M. Pereira}
\affiliation{Department of Physics, Boston University, 590
Commonwealth Avenue, Boston, MA 02215, USA}

\author{A.~H. Castro Neto}
\affiliation{Department of Physics, Boston University, 590
Commonwealth Avenue, Boston, MA 02215, USA}

\date{\today}


\begin{abstract}
We propose a route to all-graphene integrated electronic devices by exploring
the influence of strain on the electronic structure of graphene. We show
that strain can be easily tailored to generate electron beam collimation, 1D
channels, surface states and confinement, the basic elements for all-graphene
electronics. In addition this proposal has the advantage that patterning can be
made on substrates rather than on the graphene sheet, thereby protecting
the integrity of the latter.
\end{abstract}

\maketitle


%
%
%
Notwithstanding its atomic thickness, graphene sheets have been shown to
accommodate a wealth of remarkable fundamental properties, and to hold sound
prospects in the context of a new generation of electronic devices and circuitry
\cite{Geim:2007}. 
%
%
The exciting prospect about graphene is that, not only can we have extremely
good conductors, but also most active
devices made out of graphene. One of the current difficulties with respect to
this lies in that conventional electronic operations require the ability to
completely pinch-off the charge transport on demand. Although the electric field
effect is impressive in graphene \cite{Novoselov:2004}, the existence of a
minimum of conductivity poses a serious obstacle towards desirable on/off
ratios. A gapped spectrum would certainly be instrumental. The presence of a gap
is implicitly related to the problem of electron confinement, which for Dirac
fermions is not easily achievable by conventional means (like electrostatic
potential wells) \cite{Katsnelson:2006b}.
Geometrical confinement has been achieved in graphene ribbons and dots
\cite{Han:2007,Ponomarenko:2008}, but the sensitivity of transport to
the edge profile \cite{Mucciolo:2008}, and the inherent difficulty in the
fabrication of such microstructures with sharply defined edges remains a
problem.

The ultimate goal would be an all-graphene circuit. This could be
achieved by taking a graphene sheet and patterning the different
devices and leads by means of appropriate cuts that would generate leads
ribbons, dots, etc.. This
\emph{papercutting} electronics can have serious limitations with
respect to reliability, scalability, and is prone to damaging and
inducing disorder in the graphene sheet \cite{Sols:2007}. 
Therefore, in keeping with the paper art analogy, we propose an alternative
\emph{origami} electronics \cite{Tomanek:2002}.

\begin{figure}[b]
  \centering
  \subfigure[][]{
    \includegraphics[width=0.25\columnwidth]{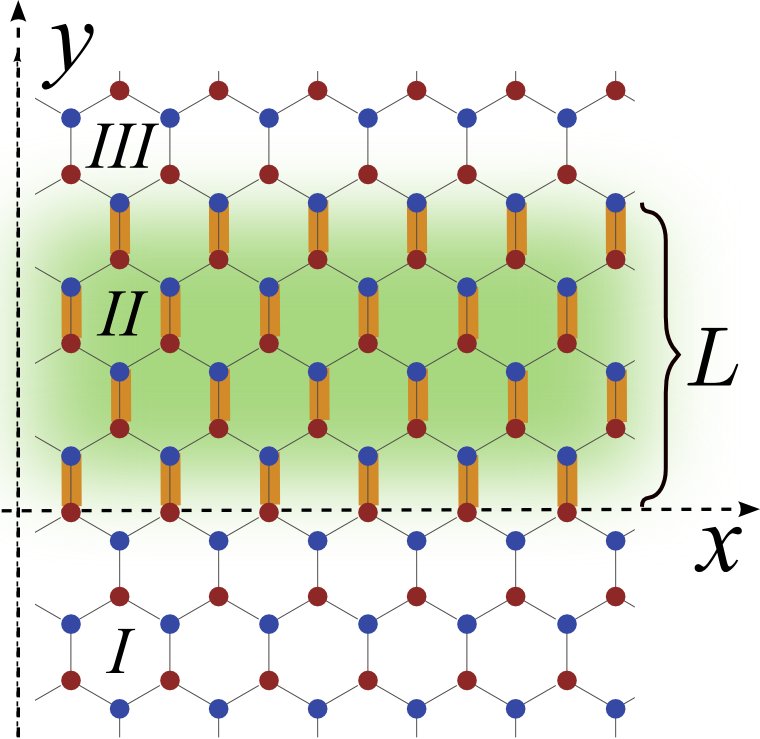}  
    \label{fig:Geometry}
  }
  \hfill
  \subfigure[][]{
      \includegraphics[width=0.6\columnwidth]{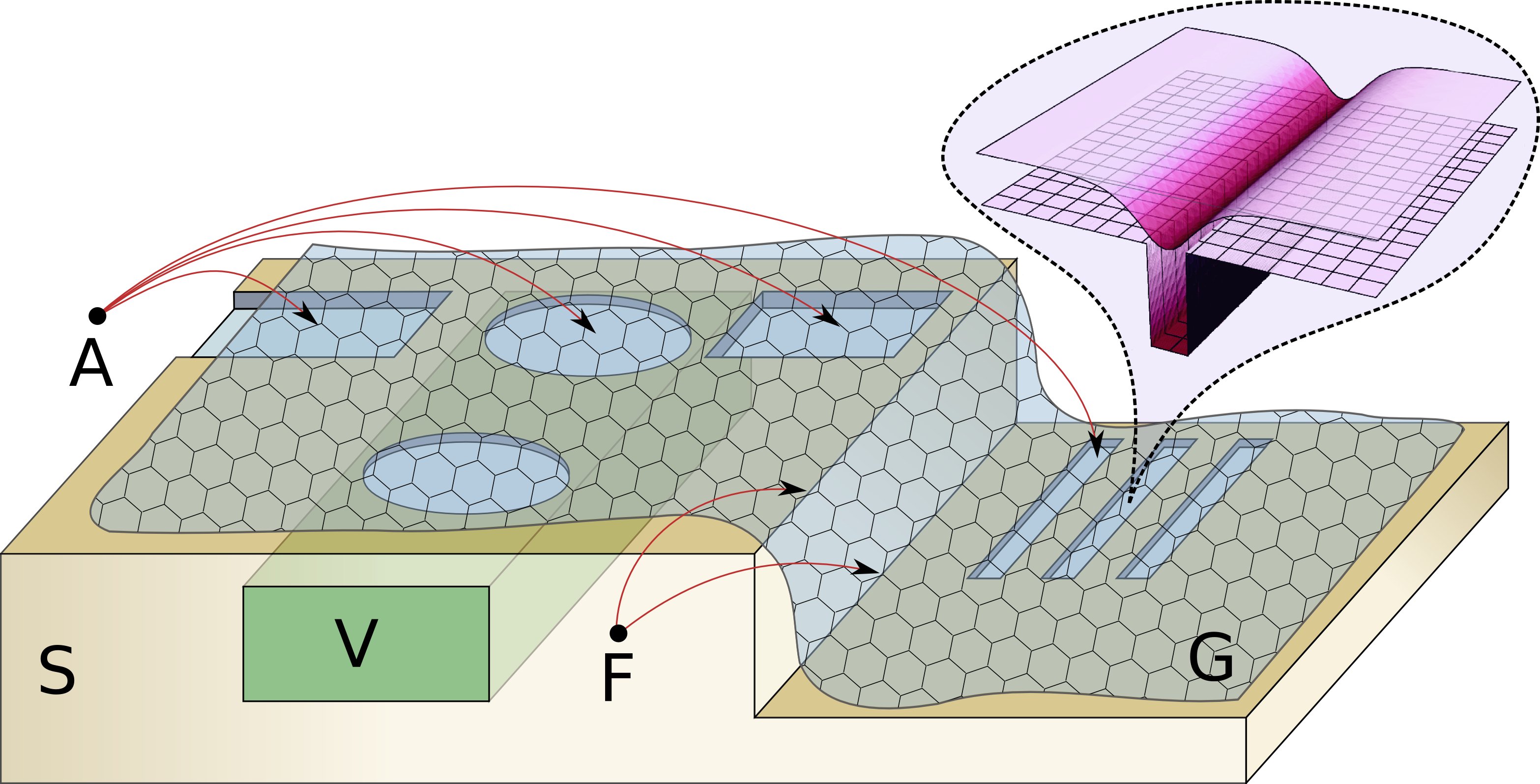}      
    \label{fig:Illustration}
  }
  \caption{
    (Color online)
    \subref{fig:Geometry} Lattice orientation considered in the text.
    Thicker bonds have perturbed hopping.
    \subref{fig:Illustration} Artistic depiction of a substrate (S)
    patterned with folds (F), trenches, dots and wells (A), upon which rests a
    graphene sheet (G). 
  }
  \label{fig:Scheme}
\end{figure}

We show here that all the characteristics of graphene ribbons and
dots (viz. geometrical quantization, 1D channels, surface
modes) might be locally obtained by patterning, \emph{not graphene}, but the
substrate on which it rests. 
The essential aspect of our approach is the generation of
strain in the graphene lattice capable of changing the in-plane hopping
amplitude in an anisotropic way. This can be achieved by means of appropriate
geometrical patterns in an homogeneous substrate (grooves, creases, steps
or wells), or by means of an heterogeneous substrate in which different regions
interact differently with the graphene sheet, generating different strain
profiles [\Fref{fig:Illustration}]. Another design alternative
consists in depositing graphene onto substrates with regions that can be
controlably strained on demand \cite{Ni:2008}.
Through a combination of folding and/or clamping a graphene sheet onto such
substrate patterns, one might generate local strain profiles suitable for the
applications discussed in detail below, while preserving a whole graphene sheet.

The remainder of the paper is dedicated to show how strain only can be
used as a means of achieving:
(i) direction dependent tunneling
(ii) beam collimation,
(iii) confinement,
(iv) the spectrum of an effective ribbon,
(v) 1D channels, and
(vi) surface modes.

%
%
%

%
%
\emph{Model --}
Within a tight-binding formulation of the electronic motion
\cite{CastroNeto:2007}, effects of in-plane strain can be captured, to leading
order, by considering the changes in nearest-neighbor hopping amplitude, $t$.
We write $t(\bR_{i},\bn) = t + \dt(\bR_{i},\bn)$, and treat the
space dependent strain-induced modulation, $\delta t$, as a perturbation
($t\approx 3\,\text{eV}$).
It is straightforward to show \cite{CastroNeto:2007} that, for smooth
perturbations, the low energy Hamiltonian is
\begin{equation}
  H = v_F\!\!\int d\br \: \Psi^\dagger
    \begin{bmatrix}
      \bm{\sigma}.(\bm{p}-\tfrac{1}{\vF}\bm{\mathcal{A}}) & 0\\
      0 & \!\!-\bm{\sigma}.(\bm{p}+\tfrac{1}{\vF}\bm{\mathcal{A}})
    \end{bmatrix}
    \Psi
  \,,
  \label{eq:H-Full}
\end{equation}
valid near the valleys $K$ and $K'$ in the Brillouin zone, with 
$\bm{\sigma}=(\sigma_x,\sigma_y,\sigma_z)$, $\vF=3ta/2\hbar$ and 
$\Psi=[\psi^A_K(\br),\,\psi^B_K(\br),\,\psi^B_{K'}(\br),\,\psi^A_{K'}(\br)]
^\dagger$ is a spinor containing the electron fields in each sublattice and
valley. Within each valley it has the form
$H=\vF\bm{\sigma}.(\bm{p}-\frac{1}{\vF}\bm{\mathcal{A}})$, so that electron
dynamics is determined by a Dirac equation in the presence of a gauge field
$\bm{\mathcal{A}}$. This field stems from the perturbation to the homogeneous
hopping amplitudes and is related to $\dt(\bR,\bn)$ via the complex
vector potential (VP) $\mathcal{A}(\br)=\mathcal{A}_x(\br)-i\mathcal{A}_y(\br)$:
\begin{equation}
  \mathcal{A}(\br) = \sum_{\bn}\dt(\br,\bn)\: e^{i\bm{K}.\bn}
\end{equation}
The fact that $\bm{\mathcal{A}}$ appears in \eqref{eq:H-Full} with its sign
reversed for the $K'$ valley guarantees overall time reversal symmetry.
For definiteness,
with the lattice orientation shown in
\Fref{fig:Geometry} we perturb the vertical hopping by a constant amount
$\dt$, over a finite region of width $L$. The perturbation and the associated
$\bm{\mathcal{A}}(\br)$ are
\begin{subequations}\label{eq:StepPerturbation}
\begin{align}
  \dt(\bR_i,\bn) &= \dt\, \delta_{\bn,0}\, \theta(Y_i)\theta(L-Y_i)
  \label{eq:StepPerturbation-a}
  \\
  \bm{\mathcal{A}}(\br) &= \dt\, \theta(y) \theta(L-y) \bu_x
  \label{eq:StepPerturbation-b}
  \,.
\end{align}
\end{subequations}
The gauge field $\bm{\mathcal{A}}$ is oriented along $\bu_x$,
which coincides with the direction of translational invariance. Using units
where $\vF=\hbar=1$, and allowing for the presence of an electrostatic
potential $V(\br)$ in the barrier region, the wave equations for the $K$
valley can then be cast as
\begin{subequations}\label{eq:WaveEqs}
\begin{align}
  \bigl[-i\partial_x-\partial_y-\mathcal{A}_x(y)\bigr]\,\psi^{B}(\br) & 
    = \bigl[E-V(\br)\bigl]\,\psi^{A}(\br)\\
  \bigl[-i\partial_x+\partial_y-\mathcal{A}_x(y)\bigr]\,\psi^{A}(\br) &
    = \bigl[E-V(\br)\bigl]\,\psi^{B}(\br)
  \,.
\end{align}
\end{subequations}
In this formulation, the problem reduces to the study of Dirac
electrons in the presence of VP and electrostatic barriers, and
is related to corresponding studies of Dirac electrons in the presence of
magnetic barriers \cite{Xu:2008,Masir:2008,Ghosh:2008,Dellanna:2008}.
The profile \eqref{eq:StepPerturbation-b} has also been considered in 
Ref.~\onlinecite{Fogler:2008} in modelling a suspended graphene sheet.

\begin{figure}
  \centering
  \subfigure[][]{
    \includegraphics*[width=0.7\columnwidth]{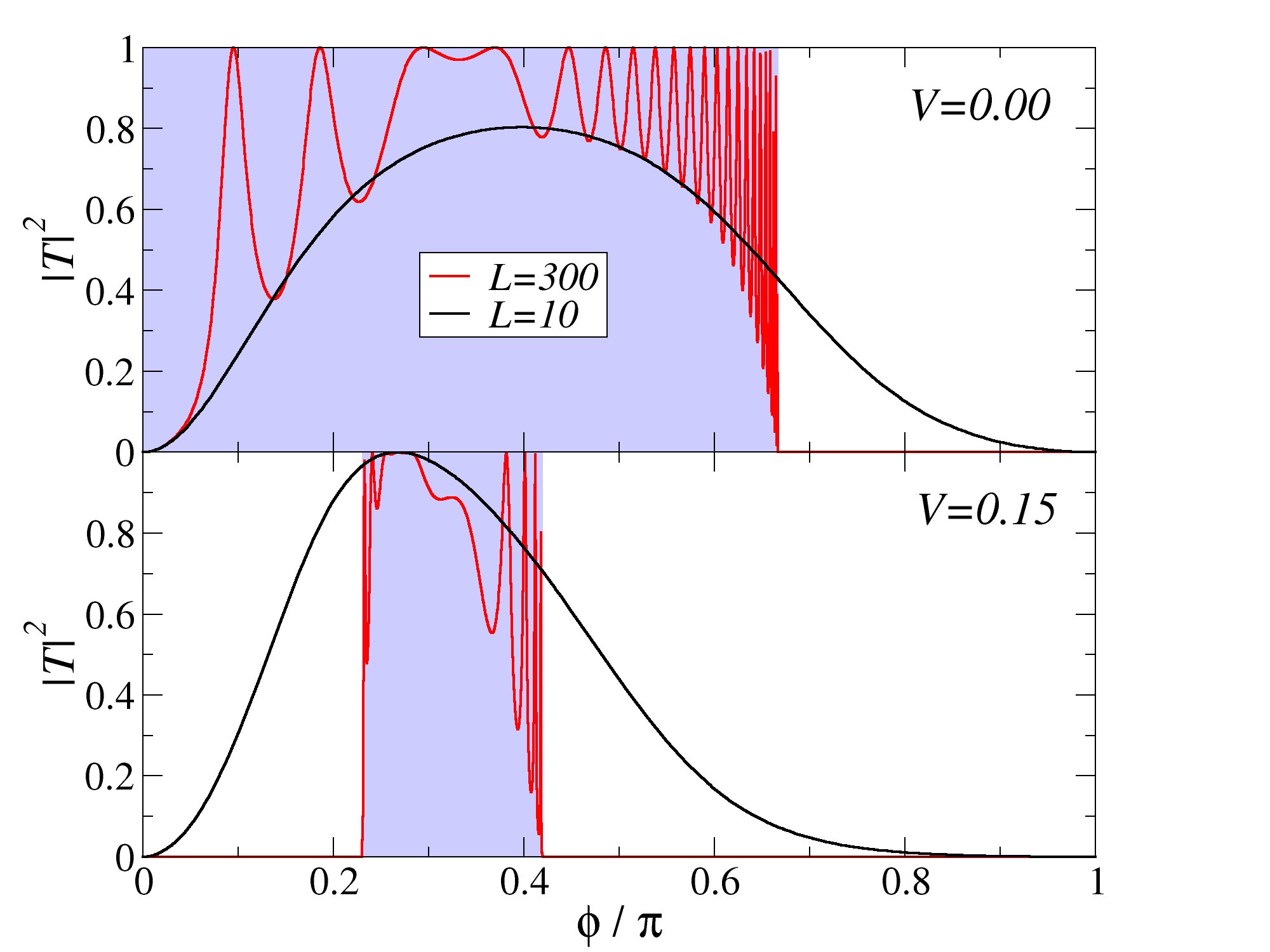}
    \label{fig:Transmission}
  }
  \subfigure[][]{
    \includegraphics*[width=0.2\columnwidth]{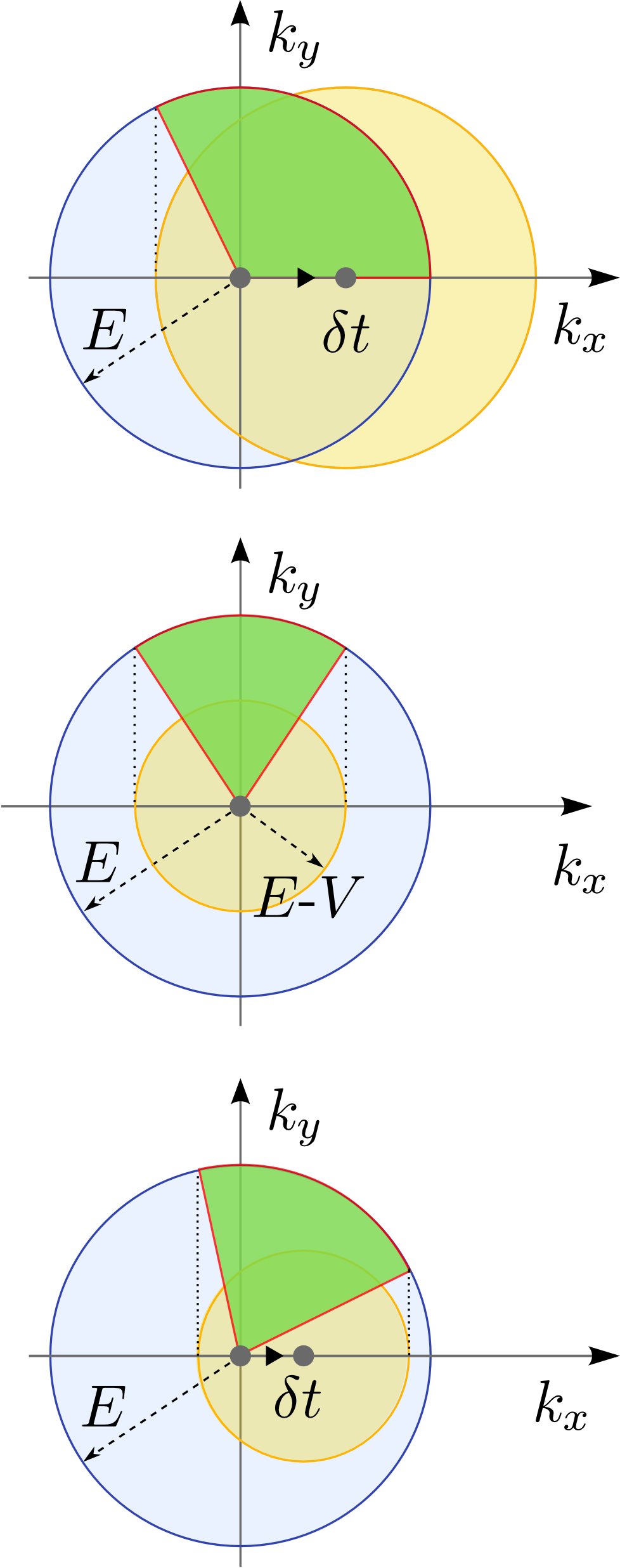}  
    \label{fig:Sectors}
  }
  \caption{
    (Color online)
    Transmission obtained from \eqref{eq:Transmission} for $L=10,\,300$, 
    $\dt= 0.1$, $E=0.2$ with (bottom) and without (top) gate potential $V$. 
    On the right we depict the phase space for the cases where
    only the strain induced VP is present $(\mathcal{A}_x\ne 0,\,V=0)$, 
    for a pure gate potential $(\mathcal{A}_x=0,\,V\ne 0)$ and
    for a combination of both $(\mathcal{A}_x\ne 0,\,V\ne 0)$.
    The shaded (green) sectors represent the range of incident angles $\phi$ 
    which are not filtered by the barrier.
  }
  \label{fig:Tunneling}
\end{figure}

%
%
\emph{Tunneling --}
We begin by analyzing the tunneling characteristics across the barrier-like
perturbation of \Eqref{eq:StepPerturbation}. 
Without compromising generality \cite{Endnote-1}, in the remainder of the paper
we shall be concerned with the situation $\dt>0$ and $E>0$. 
We parametrize the wavefunction in the three regions of \Fref{fig:Geometry} as
\begin{eqnarray*}
  \Psi^I & = & e^{ik_x x+ik_y y}\,
    \binom{1}{e^{i\phi}}
    +R\,e^{ik_{x}x-ik_{y}y}\,
    \binom{1}{e^{-i\phi}}\\
  \Psi^{II} & = & C_1\binom{1}{e^{i\varphi}}
    e^{ik_x x+iq_y y}+C_{2}\binom{1}{e^{-i\varphi}}
    e^{ik_x x-iq_y y}\\
  \Psi^{III} & = & T\, e^{ik_x x+ik_y y}\,
    \binom{1}{e^{i\phi}}
\end{eqnarray*}
where we have the kinetic momenta $k_x=E \cos\phi$, $k_x-\dt=E
\cos\varphi$, $q_y=E\sin\varphi$ and the energy $E^2=k_x^2+k_y^2=(k_x-\delta
t)^2+q_y^2$. 
Substitution in the wave equation \eqref{eq:WaveEqs} leads to the
following transmission amplitude \cite{Ghosh:2008,Fogler:2008}:
\begin{equation}
  T\!=\! \frac{  e^{-ik_y L}\sin\phi\sin\varphi }{
    \cos(q_y L)\sin\phi\sin\varphi\!+\!i\sin(q_y L)(\cos\phi\cos\varphi\!-\!1)
  }.
  \label{eq:Transmission}
\end{equation}
This result is valid also for $V\ne 0$ with the appropriate substitution
$E\to E-V$ inside the barrier. 
As pointed out in the case of a real magnetic field \cite{Martino:2007},
conservation of $k_x$ requires $E \cos\phi = \dt + E \cos\varphi$, leading
to strong suppression of tunneling for $\phi>\arccos(-1+\dt/E)$. When
such condition is in effect, the internal angle has to be analytically
continued to the imaginary axis $\varphi\to i\varphi\,,q_y\to iq_y$ causing
an exponential suppression of $|T|^2$. Moreover, if $\dt/E>2$ tunneling
is completely suppressed. These effects are illustrated in
\Fref{fig:Transmission} where we plot $|T|^2$ for different values of $L$ and
$V$.
%
%
%

%
%
\emph{Beam Collimation --}
Several aspects can be immediately identified from Fig. \ref{fig:Tunneling}. 
Klein tunneling \cite{Katsnelson:2006b} is absent for a pure VP barrier, and if
$V\ne0$ Klein tunneling might persist albeit at $\phi\ne\pi/2$. As hinted above,
for wide barriers tunneling is highly suppressed for certain ranges of $\phi$
that depend on $\dt$, $E$ and $V$. This filtering effect for certain
incidence angles is best appreciated by inspecting the phase space
pictures shown in \Fref{fig:Sectors}: the effect of $\mathcal{A}_x$ is to
translate the Fermi surface $E^2=k_x^2+k_y^2$ by $\dt$ along the
horizontal axis. Conservation of energy and momentum immediately leads to a
sector of allowed incident angles, as drawn. Analogous reasoning applies for a
pure electrostatic barrier (where the Fermi surface in the barrier changes
size) or a combination of both. In all cases this geometrical construction
immediately yields the transmission sector. It is clear that, whereas in a
purely electrostatic barrier the transmission sector is symmetric with respect
to normal incidence ($\phi=\pi/2$), in a VP barrier this sector always contains
either $\phi=0$ or $\pi$, depending on the relative sign of $E$ and $\dt$.
Combining these two cases one can generate virtually any transmition sector with
a single barrier, as exemplified in \Fref{fig:Tunneling}.

This has immediate applicability in electron beam collimation and lensing, and 
the effect is easily amplifiable through a series of barriers \cite{Ghosh:2008}.
In addition, the beam refracted by the barrier can approach or recede from
the normal interchangeably by changing the sign of $\dt$. 
Alternatively, since the transmission sector depends explicitly on $E$,
a suitable geometrical configuration of barriers can be used to filter the 
energy of the incoming beam.
In addition, the suppression of tunnelling for certain angular
sectors leads to the appearance of a transport gap, as shown in
Ref.~\onlinecite{Fogler:2008}. Therefore, even though small strain does not
lead to a bulk spectral gap, the system exhibits an effective transport gap.
Within the allowed sector, transmission through wider barriers is additionally
characterized by a series of marked resonances where $|T|^2=1$, and electron
flow is totally unhindered. Such behavior, usually associated with successive
internal reflections, strongly suggests the possibility of confinement.

%
%
\begin{figure}[tb]
  \centering
  \subfigure[][]{%
    \includegraphics*[width=0.52\columnwidth]{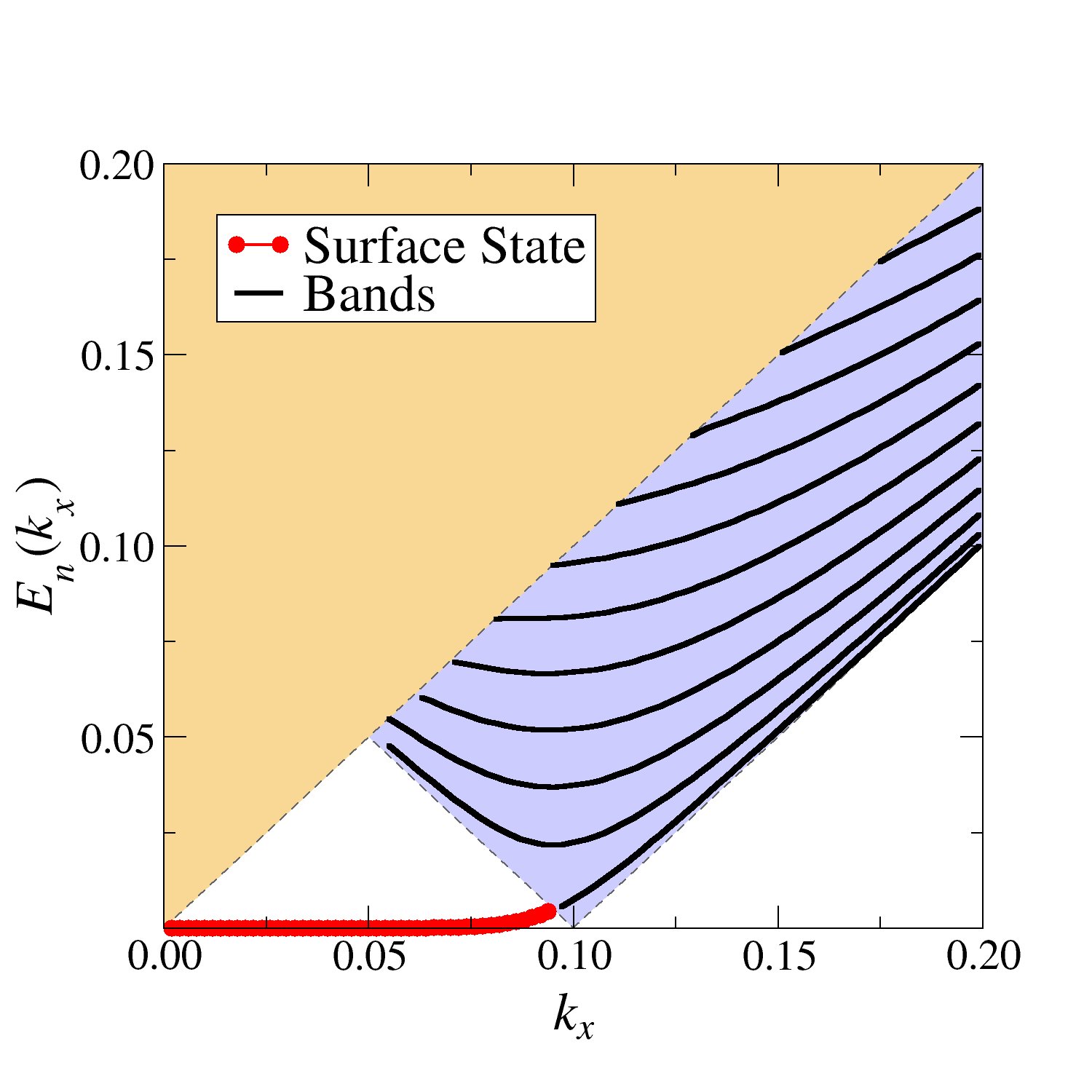}
    \label{fig:ConfinedStates}
  }%
  \subfigure[][]{%
   
\includegraphics*[width=0.48\columnwidth]{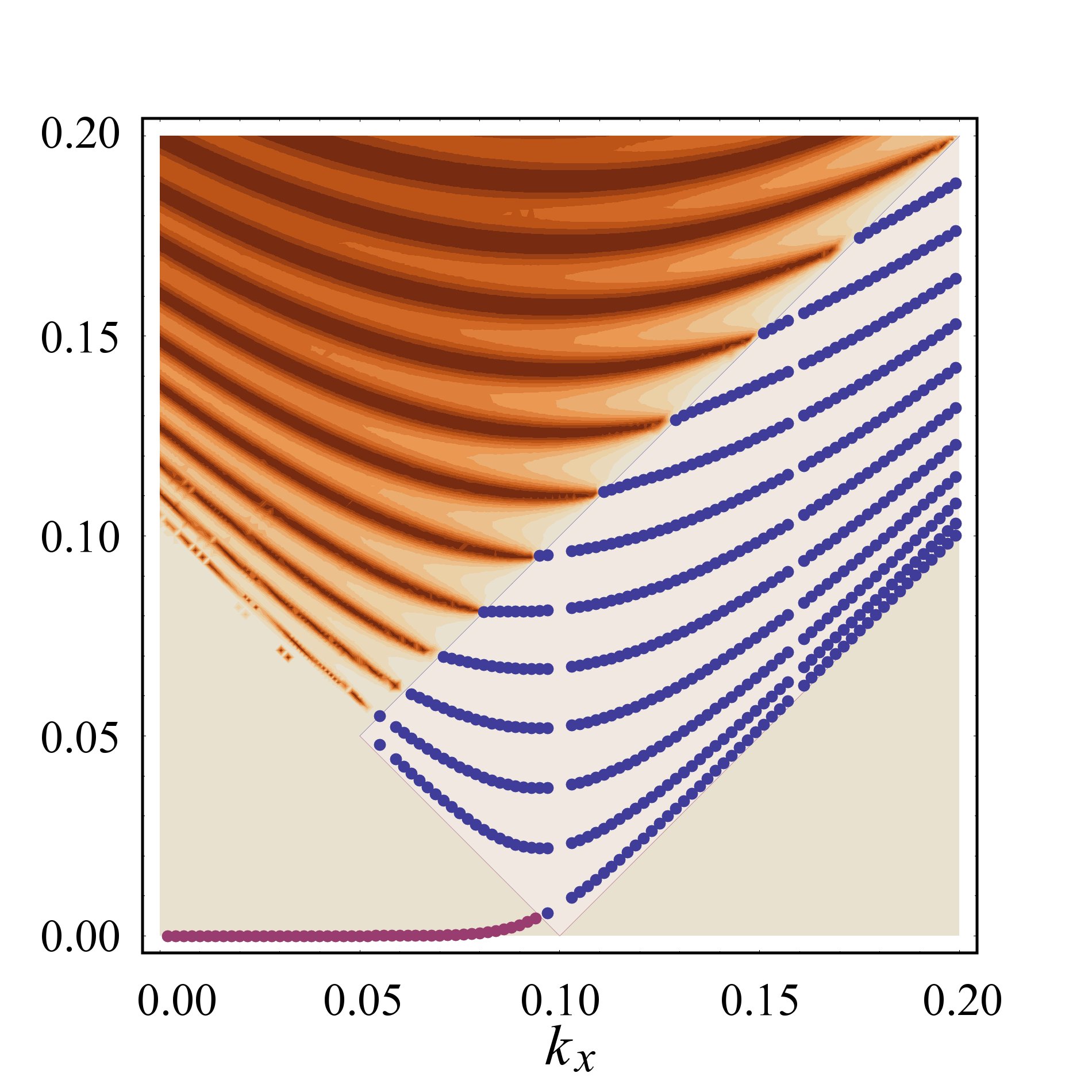}
    \label{fig:ConfinementTogether}
  }
  \subfigure[][]{%
    \includegraphics*[width=0.38\columnwidth]{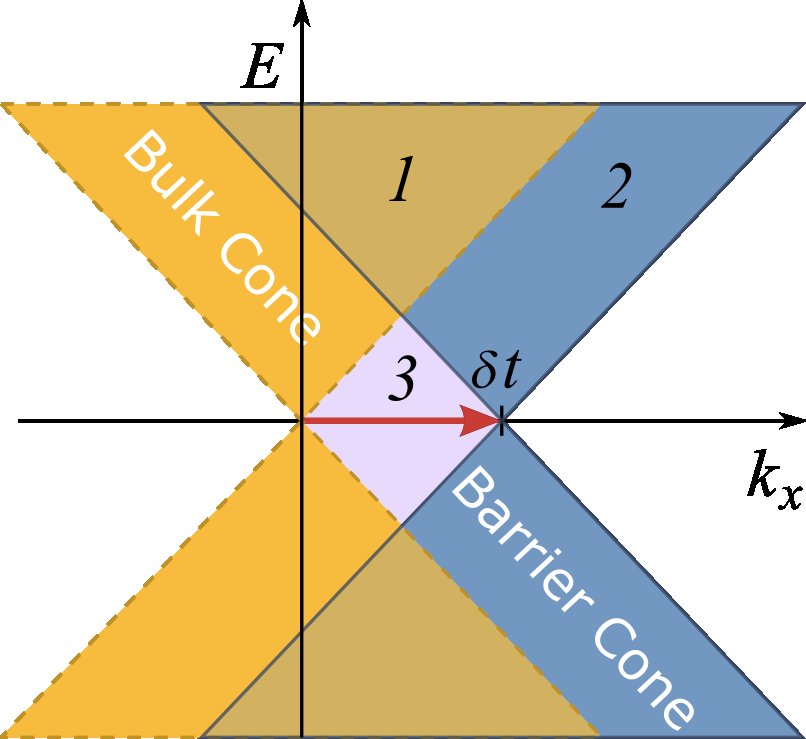}
    \label{fig:Cones}
  }
  \qquad
  \subfigure[][]{%
    \includegraphics*[width=0.45\columnwidth]{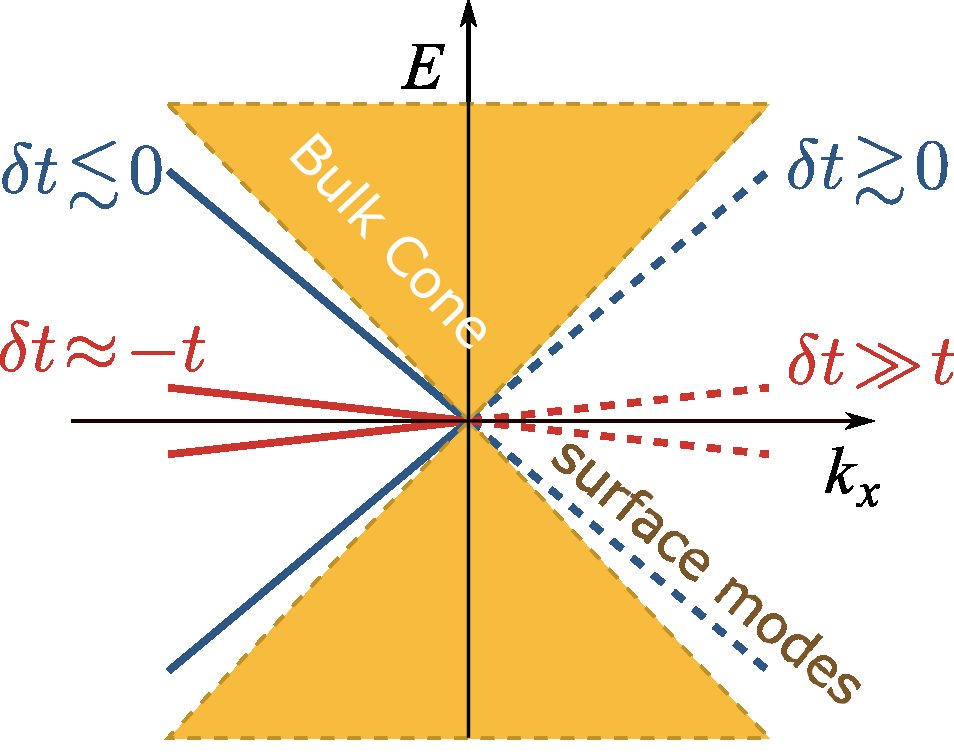}
    \label{fig:Narrow}
  }
  \caption{
    (Color online)
    \subref{fig:ConfinedStates} Dispersion of the confined solutions,
      $E_n(k_x)$ for $\dt=0.1$ and $L=200$.
    \subref{fig:ConfinementTogether} A density plot of the transmission
      $t(E,k_x)$ from \Eqref{eq:Transmission} is shown in the upper
      region (region 1), which is plotted together with the same data as in 
      panel \subref{fig:ConfinedStates}.
    \subref{fig:Cones} Regions 1, 2 and 3 discussed in the text.
    \subref{fig:Narrow} Linearly dispersing modes for the narrow barrier 
      discussed in the text.
  }
  \label{fig:Confinement}
\end{figure}
%

%
%
\emph{Confinement --}
Dirac electrons are notoriously resilient to conventional confinement
strategies on account of the Klein paradox \cite{Katsnelson:2006b}. The fact
that our VP barriers exponentially suppress electronic transmission will be
further explored to confine Dirac electrons. A confined state inside the barrier
has the form
\begin{gather}
  \Psi^{I} = A\, e^{ik_{x}x}e^{\kappa y}\,\binom{1}{e^\vartheta}
  ,\,
  \Psi^{III} =D\, e^{ik_{x}x}e^{-\kappa y}\,\binom{1}{e^{-\vartheta}}
  ,\nonumber\\
  \Psi^{II} =B\, e^{ik_{x}x}e^{ik_y y}\binom{1}{e^{i\varphi}}
    +C\,e^{ik_x x}e^{-ik_y y}\,\binom{1}{e^{-i\varphi}}
  \,.
  \label{eq:WF-2}
\end{gather}
From the wave equation it follows that 
$E^2=k_x^2-\kappa^2=(k_x-\dt)^2+k_y^2$, $k_x=E \cosh\vartheta$, 
$\kappa = E \sinh\vartheta$, $k_x-\dt=E \cos\varphi$, $k_y=E \sin\varphi$. From
the fact that $E^2>0$, this type of solution requires $|k_x-\dt|<|E|<|k_x|$.
This is graphically represented by region 2 in \Fref{fig:Cones}. In addition,
continuity of the wavefunction requires that
\begin{equation}
  \cot(k_y L)\sin\varphi\sinh\vartheta=1-\cos\varphi\cosh\vartheta
  \label{eq:quantization}
\end{equation}
be satisfied. When solved for $E$, \Eqref{eq:quantization} yields a
discrete spectrum of energies for each value of $k_x$: $E_n(k_x)$. Region 2
shown in \Fref{fig:Cones} is therefore characterized by the emergence of 
\emph{1D channels (quantum wire) confined to the barrier region}, and dispersing along $k_x$. A
particular realization of this effect is shown in \Fref{fig:ConfinedStates}. It
is clear from this figure and Eqs.~(\ref{eq:WF-2},\ref{eq:quantization}) that
such states share all the features of the 1D modes typical of
graphene nanoribbons. In particular, at the displaced Dirac point the
spectrum scales as 
$E_n(k_x=\dt)\approx (n+1/2)\pi/L$ and, for all
purposes related to the electronic states and spectrum, in region 2 this system
behaves as a nanoribbon. This includes the energy scales associated with the
confinement gaps. For example, the gap is roughly 
$E_g \approx 1/L$~eV.nm, so that $L\approx20$nm yields a gap of $\approx50$~meV.
This is valid as long as $\delta t\gg E_g$, which should be the case for $\delta
t/t\sim 10\%$. Hopping variations of this magnitude can be achieved with a local
strain of $\sim 5\%$ \cite{Pereira:Strain}.

%
%
\emph{Surface Modes --}
The similarity of a VP barrier with the physics of a nanoribbon achieves its
fullest if one notices that surface states are also possible. Just like the edge
modes of zig-zag nanoribbons, we consider a state localized at the barrier
edges. Such state should decay to both sides of $y=0$ and $y=L$ simultaneously.
Consequently we can construct its wavefunction from \Eqref{eq:WF-2} through
analytical continuation of $k_y$ to the imaginary axis: $k_x\to
i\kappa'$. The energy will be $E^2=k_x^2-\kappa^2=(k_x-\dt)^2-\kappa'^2$, which
restricts the space of solutions to the areas outside both Dirac cones
[\Fref{fig:Cones}]. The solution for the wavefunction so constructed leads to
a quantization condition analogous to \eqref{eq:quantization} but, since
the circular functions are converted to hyperbolic, will admit only
one solution, valid precisely when $0<k_x\le\dt$ [region 3 of
\Fref{fig:Cones}]. In region 3 we then have a single state, whose dispersion is
shown in \Fref{fig:ConfinedStates}. 
This mode smoothly merges with the lowest state in region 2 and, in the limit
of $L\to\infty$ its energy is given by 
$E\approx2(\dt-k_x)\sqrt{\frac{k_x}{\dt}}\, e^{-L(\dt-k_x)}$, 
making clear that this mode's energy decays exponentially from the shifted
Dirac point towards $k_x=0$. 
Inspection of the solution in this limit reveals that the amplitude in one 
sublattice becomes much smaller than in the
other, in complete analogy to the edge states in a wide nanoribbon.
It is straightforward to demonstrate that such surface states persist in the
limit $L=\infty$ where the problem reduces to a VP step. For a step, the surface
modes have strictly zero energy, and occupy only one sublattice
\cite{Endnote-2}.

The results shown in \Fref{fig:ConfinedStates} are symmetric with respect to
$E=0$ \cite{Endnote-1}. This figure shows that the region $2\cup 3$, lying
outside the continuum of the bulk of the system, supports spatially confined
1D modes, which have the same characteristics as modes in a ribbon, including
the presence of surface states. We can now coherently
interpret what happens in region 1, in connection with the
tunneling calculations described earlier:
just as in a conventional Schr\"odinger-like barrier, the 
scattering states in region 1 (the bulk continuum) should ``feel'' the presence
of the confined solutions inside the barrier. This manifests itself
through the transmission resonances already shown in \Fref{fig:Transmission}.
In fact, when the transmission \eqref{eq:Transmission} is plotted in the
$(E,k_x)$ plane, we obtain the result shown in
region 1 of \Fref{fig:ConfinementTogether}. The darker regions of this 
density plot correspond to the transmission resonances, which lie
in the extrapolation of the confined modes into the bulk continuum.

%
%
\emph{Narrow Limit --}
The approach used so far begs revision for the case of very narrow and/or
large perturbations in the hopping. 
The clearest example is the case where $L$ approaches the lattice spacing, so
that hopping is perturbed only in one unit cell along $\bu_y$. Clearly, if
$\dt=-t$ the upper and lower half planes become disconnected and no tunneling is
expected. This, however, is not captured by \Eqref{eq:Transmission} which
assumes overall continuity of the wavefunction and small perturbations. Since
the narrow limit is of interest, e.g. in the case of a tight graphene crease, we
tackle it by analyzing the localized perturbation
\begin{equation}
  \dt(\bR_i,\bn) = \dt\, \delta_{\bn,0} \delta_{Y_i,0}
  \Rightarrow
  \bm{\mathcal{A}}(\br) = \hbar\vF\tfrac{\dt}{t}\, \delta(y) \bu_x
  \label{eq:DeltaPerturbation}
  \,.
\end{equation}
One consequence of the first order nature of the Dirac equation
\eqref{eq:WaveEqs} is the discontinuity of the wavefunction in the presence of
this potential. The presence of $\delta(y)$ in
\eqref{eq:DeltaPerturbation} imposes a boundary condition at the origin. In our
case, there is the peculiarity that the weight of $\delta(y)$ cannot be even for
both sublattices: for the geometry of \Fref{fig:Geometry} the
appropriate integration of this Dirac-delta is
\begin{equation}
  \int_{0^-}^{0^+}dy\,\psi^{A,B}(x,y)\, \delta(y) =
\psi^{A,B}(x,0^{+,-})
  \label{eq:DiracDelta}
  \,
\end{equation}
leading to the boundary conditions
$  \psi^B(0^-)=\eta\, \psi^B(0^+)$
and 
$  \psi^A(0^+)=\eta\, \psi^A(0^-) $,
with $\eta=1+\dt/t$. The wavefunction is naturally discontinuous and, moreover,
for $\dt= -t$ ($\eta=0$) vanishes in different sublattices at the upper and
lower half-plane. Notice that in the case $\dt= -t$ the system is comprised of
two independent semi-planes with a zig-zag termination. Our BC reproduces the
correct BC for these two zig-zag
semi-planes \cite{CastroNeto:2007,Endnote-3}. The transmission and reflection
amplitudes as a
function of $\phi=\arg(k_x+ik_i)$ read
\begin{equation}
  R = \frac{1-\eta^2}{\eta^2-e^{-2i\phi}}
  \,,\quad
  T = \eta \, \frac{1-e^{-2i\phi}}{\eta^2-e^{-2i\phi}}
  \label{eq:r-t-narrow}
  \,,
\end{equation}
with $|R|^2+|T|^2=1$, and $T=0$ for $\eta=0$ as expected. The interesting fact
about such narrow barriers is that \emph{they still support surface modes}.
Applying the same procedure outlined above to the case
\eqref{eq:DeltaPerturbation} one straightforwardly obtains a state
that decays exponentially to both sides of the axis $y=0$. Its dispersion is
given by 
$ E(k_x)=\pm v_s |k_x| $ where $v_s = 2|\eta|/(1+\eta^2)$
and such states exist for $k_x>0\, (k_x<0)$ when $\eta^2>1\, (\eta^2<1)$. This
is a very interesting situation: the perturbation of a single row of hoppings
leads to the emergence of linearly-dispersing 1D modes that live along the
perturbed line. These modes detach from the continuum for small $\dt$
($E\lesssim\pm|k_x|$), and reach zero energy at $\dt=-t$, just as expected for
the two zig-zag semiplanes that result in that limit. This effect is
illustrated in \Fref{fig:Narrow}.

In summary, our approach demonstrates that strain-induced gauge fields can be
tailored to generate confined states, quantum wires and collimation in graphene.
These results, together with the fact that strain has been reliably controlled
in graphene \cite{Ni:2008}, opens an exciting prospect towards all-graphene
electronics.

%
%
\acknowledgments
We acknowledge the hospitality of the Aspen Center for Physics,
where this work germinated.
VMP is supported by FCT via SFRH/BPD/27182/2006 and 
PTDC/FIS/64404/2006. 
AHCN was partially supported by the U.S. DOE
under the grant DE-FG02-08ER46512.


\bibliographystyle{apsrev}
\bibliography{graphene_origami}

\begin{thebibliography}{20}
\bibitem[{End({\natexlab{a}})}]{Endnote-1}
\bibinfo{note}{Our results are symmetric with respect to $E\to-E$. When
  $\dt\to-\dt$ the Dirac cone is shifted in the opposite direction and thus our
  results should be reflected along the axis $k_x=0$.}
\bibitem[{End({\natexlab{b}})}]{Endnote-2}
\bibinfo{note}{The sign of $\dt$ in the step determines which sublattice.}
\bibitem[{End({\natexlab{c}})}]{Endnote-3}
\bibinfo{note}{For the orientation of Fig.~\ref{fig:Geometry} there are no
  intervalley terms, just as in the case of a pure zig-zag edge.}
\end{thebibliography}

\end{document}